\documentstyle[12pt]{article}

\def\tr{\mathop{\rm tr}\nolimits}

\begin{document}
\begin{titlepage}
\begin{flushright}
{\bf SPbU-IP-94-17}
\end{flushright}
\vskip 2cm
\begin{center}
{\LARGE Extended Chiral Transformations Including \\[8pt]
Diquark Fields As Parameters}
\vskip 2cm
{\bf Yuri Novozhilov, Andrei Pronko and Dmitri Vassilevich}
\vskip 1cm
{\it  Department of Theoretical Physics\\
Research Institute for Physics\\
St.~Petersburg State University\\
198904 St.~Petersburg\\
RUSSIA}
\end{center}
\vskip 5cm
\begin{abstract}
\normalsize

We introduce extended chiral  transformation, which depends both
on pseudoscalar and diquark fields as parameters and determine its group
structure. Assuming soft symmetry breaking in diquark sector, bosonisation
of a quasi-Goldstone $ud$-diquark is performed. In the chiral limit the
$ud$-diquark mass is defined by the gluon condensate,
$m_{ud}\approx 300 MeV$. The diquark charge radius is
$\langle r^2_{ud}\rangle^{1/2}\approx 0.5 fm$.

\end{abstract}
\end{titlepage}

\section{Introduction.}

Diquarks, which were introduced almost three decades ago [1],
became now efficient tool for studying various processes in hadron physics
(see e.g. [2,3] and reviews [4]). The diquark model was analysed from
various points of view. However, the complete picture is still lacking.

It was suggested by Dosch et al [3] that wave function of pion and
$ud$-diquark  are the same at the origin. It was also shown [3] that the
diquark decay constant following from this suggestion is very close to that
estimated from the QCD sum rules.

In this paper we propose to go a bit further: we suppose that the
similarity of wave functions of pion and diquark  is due to common origin
as parameters of a certain anomalous transformation which does not
preserve the measure of the quark path integral. While at the classical
level the chiral symmetry is broken by quark mass, the extended chiral
(E$\chi$) symmetry is broken by quark mass and gluon fields. E$\chi$-group
is $U(2N)$ for $N$ internal degrees of freedom, $N=N_c N_f$. Non-anomalous
(measure preserving) generators span the Lie algebra of $O(2N)$, anomalous
generators belong to the coset $U(2N)/O(2N)$. Anomalous generators describe
chiral rotations and transformations with diquark variables (``diquark''
rotations), non-anomalous part consists out of gauge transformations and
combined chiral ``diquark'' rotations.

We assume that E$\chi$-symmetry breaking due to quark masses and gluon
fields is soft in the sense that the action for bosonised diquark fields
can be obtained by integrating corresponding E$\chi$-anomaly. Colorless
chiral fields after bosonisation give rise to Goldstone particles --
pseudoscalar mesons. We suggest that at low energies bosonised diquark
parameters of E$\chi$-transformations with quantum numbers of lightest
$J^P=0^+$ $ud$-diquark can be treated as a Goldstone-like particle.
Therefore, in bosonisation we restrict ourselves  to the case of
E$\chi$-transformations
with $ud$-diquark fields. The E$\chi$-group in this case is
$SU(4)$, non-anomalous transformations are just gauge transformations
$SU(3)\times U(1)$ and the diquark Goldstone degrees of freedom belongs to
$CP^3=SU(4)/SU(3)\times U(1)$.

Our analysis shows that the $ud$-diquark introduced
{\it a la} Goldstone becomes massless in the limit of vanishing
gluon condensate and current quark masses. Furthermore, we calculate the
diquark mass and charge radius for the actual  value of gluon condensate.
The obtained values fall into the region allowed  in other models [4].
Note, that our approach is a direct generalization of chiral bosonisation
scheme [5] for the case of new anomalous transformation. We introduce no
new parameters.

\section{Group  structure of E$\chi$-transformations.}

In our previous paper [6] we demonstrated that in order to consider
quark-antiquark and quark-quark composites on equal footings one should
introduce eight-component spinors $\Psi$ constructed from ordinary Dirac
spinors $\psi$
\begin{equation} \label{Psi}
\Psi=\left( \begin{array}{l} \psi \\
                            \bar\psi^T
\end{array} \right)
\end{equation}
The quark lagrangian can be rewritten in the form
\begin{equation} \label{lag}
{\cal L} = \frac{1}{2} \Psi^T \hat F \Psi,\qquad
\hat F=\left(
\begin{array}{cc}  C\Phi & -D^T \\
				   D     &  \overline{\Phi}C
\end{array} \right) \qquad
F=-F^T
\end{equation}
where $D$ is the Dirac operator
$D=i\gamma^\mu(\partial_\mu+v_\mu+\gamma_5 a_\mu)$, ``${}^T$'' means
transposition and $\Phi=\gamma^\mu(\phi_{5\mu}+\gamma_5\phi_\mu)$,
$\overline\Phi=\gamma_0\Phi^+\gamma_0$. We have introduced various
external fields $v_\mu,a_\mu,\phi_\mu$ and $\phi_{5\mu}$ generating both
$\bar\psi \psi$ and $\psi\psi$ composites. $C$ is charge conjugation
matrix. The quark path integral becomes
\begin{eqnarray} \label{Z-psi}
Z_\psi & = & \int {\cal D}\Psi \exp i\int d^4x {\cal L} = (\det\hat G)^{1/2}
\nonumber\\
\hat G &=& \left(\begin{array}{cc} D & \overline{\Phi}\\
							\Phi   & D_c
\end{array}\right), \qquad
D_c=C^{-1}D^T C,
\nonumber\\
\hat F &=& \left(\begin{array}{cc}
0 & -1\\
1 &  0
\end{array}\right)
\left(\begin{array}{cc}
1 & 0 \\
0 & C^{-1}
\end{array}\right)
\hat G
\left(\begin{array}{cc}
1 & 0\\
0 & C
\end{array}\right)
\end{eqnarray}
where the operator $\hat G$ is $\gamma_0$-hermitian.
A similar construction was considered by Ball [7] for Majorana spinors.

The lagrangian (\ref{lag}) is invariant under the following
transformations
\begin{equation} \label{Omega}
\delta\Psi=-\Omega\Psi, \qquad
\Omega=
\left(\begin{array}{cc}
\alpha+\gamma_5\chi & (\xi+\gamma_5\omega)C\\
(-\xi^*+\gamma_5\omega^*)C & \alpha^*-\gamma_5\chi^*
\end{array}\right)
\end{equation}
provided  external fields in the operators $\hat G$ and $\hat F$
transform according to the rules
\begin{eqnarray} \label{Xi-Theta}
\hat F \rightarrow  \hat F' & = & \exp\Omega^T \, \hat F \, \exp\Omega
\nonumber\\
\hat G  \rightarrow \hat G' & = & \exp(-\Xi+\gamma_5\Theta)\, \hat G  \,
									   \exp(\Xi+\gamma_5\Theta)
\nonumber\\
\Xi & = & \left(\begin{array}{rr}
\alpha   &  \xi  \\
\xi^*   &  \alpha^*
\end{array}\right)
\qquad
\Theta=\left(\begin{array}{rr}
\chi      & \omega \\
-\omega^* & -\chi^*
\end{array}\right).
\end{eqnarray}
The matrices  $\alpha$ and $\chi$ are antihermitian, $\xi$ is antisymmetric
and $\omega$ is symmetric in internal indices. The transformations
(\ref{Omega}) do not destroy the structure (\ref{Psi}) of the
eight-component spinor $\Psi$. These transformations can be absorbed in
transformations of background fields.

Due to the noninvariance of the measure, only part of the transformations
(\ref{Omega}) do not change the path integral (\ref{Z-psi}).
These are the transformations generated by $\Xi$. The generators $\Theta$
lead to quantum anomalies. The operators $\alpha$ generate gauge
transformations, $\chi$ describe chiral rotations, the anomalous
transformations $\omega$ include fields with  diquark quantum numbers,
the generators $\xi$ are needed for closure of the algebra.

The matrix commutator
\begin{equation} \label{com}
[\Xi(\alpha_1,\chi_1)+\gamma_5 \Theta(\xi_1,\omega_1),
\Xi(\alpha_2,\chi_2)+\gamma_5 \Theta(\xi_2,\omega_2)]
=\Xi(\alpha_3,\chi_3)+\gamma_5 \Theta(\xi_3,\omega_3)
\end{equation}
induces the following Lie structure
\begin{eqnarray} \label{laws}
\alpha_3&=&[\alpha_1,\alpha_2]+[\chi_1,\chi_2]+\xi_1\xi_2^*
-\xi_2\xi_1^*-\omega_1\omega_2^*+\omega_2\omega_1^*,
\nonumber\\
\chi_3&=&[\alpha_1,\chi_2]-[\alpha_2,\chi_1]-\xi_1\omega_2^*
-\omega_2\xi_1^*+\xi_2\omega_1^*+\omega_1\xi_2^*,
\nonumber\\
\xi_3&=&\alpha_1\xi_2+\xi_1\alpha_2^*-\alpha_2\xi_1-\xi_2\alpha_1^*
+\chi_1\omega_2-\omega_1\chi_2^*-\chi_2\omega_1+\omega_2\chi_1^*,
\nonumber\\
\omega_3&=&\alpha_1\omega_2+\chi_1\xi_2-\xi_1\chi_2^*+\omega_1\alpha_2^*
-\alpha_2\omega_1-\chi_2\xi_1+\xi_2\omega_1^*-\omega_2\xi_1^*.
\end{eqnarray}
One can verify that the composition laws (\ref{laws}) are induced also
by the
matrix commutator without $\gamma_5$
\begin{equation} \label{out-5}
[\Xi(\alpha_1,\chi_1)+\Theta(\xi_1,\omega_1),
\Xi(\alpha_2,\chi_2)+ \Theta(\xi_2,\omega_2)]
=\Xi(\alpha_3,\chi_3)+\Theta(\xi_3,\omega_3)
\end{equation}
This means that the Lie algebras (\ref{com}) and (\ref{out-5}) are
isomorphic. For the case of $N$ internal degrees of freedom, $N=N_c N_f$,
and maximally extended algebra (i.e. when $\alpha, \chi, \xi$ and $\omega$
are all matrixes satisfying the above mentioned hermiticity and symmetry
properties), the $\Xi+\Omega$ form the space of hermitian matrices
$2N\times 2N$. Hence the algebra (\ref{out-5}) is $U(2N)$. The generators
$\Xi$ preserve symmetric non-degenerate bilinear form $O$
\begin{equation} 
O=\left(
\begin{array}{cc}
0 & 1 \\
1 & 0
\end{array}
\right), \qquad \Xi O + O \Xi^T=0.
\end{equation}
Consequently the non-anomalous generators $\Xi$ span the Lie algebra of
$O(2N)$ and the anomalous generators  belong to the coset $U(2N)/O(2N)$.
The generators $\alpha $ and $\omega$ preserve symplectic form
\begin{equation} 
  \Sigma=\left(
\begin{array}{cc}
0 & 1 \\
-1 & 0
\end{array}
\right)
\end{equation}
and  thus span the Lie algebra of the subgroup $Sp(N)$. The generators
obviously form $U(N)$.

In principle, any transformation of $\Theta$ could be related to a Goldstone
particle, whose dynamics is governed by quantum anomalies. However in
realistic models most of the symmetries (\ref{Omega}) are broken already at
classical level by the presence of  quark masses and gluon fields, and
the vector field prescribed by the transformation rules (\ref{Xi-Theta}).
Only colorless chiral fields $\chi$ are definitely interpreted as
pseudoscalar mesons. Some other states were also considered in literature
[8]. We suggest, that at certain energy scale the fields $\omega$ with
quantum numbers of lightest $J^P=0^+$ $ud$-diquarks can also generate
Goldstone-like particles. In what follows we shall restrict ourselves to the
transformations
\begin{equation} \label{omega}
\omega=(1/F_\omega)\omega_c(i\sigma_2)_{jk}\epsilon_{abc}
\end{equation}
corresponding to $0^+$ $ud$-diquarks where j,k are flavor  and a,b,c are
color indices.

In this special case transformations close in a smaller group. To see this
one should exclude the $i\sigma_2$ in the same way, as it was done
previously with $\gamma_5$, and use the commutation relations
(\ref{laws}). One can obtain that after removing $i\sigma_2$ and $\gamma_5$
the algebra becomes  formally equivalent to that generated by the $\alpha$
and $\xi$ operators in the case $N=3$. Hence, the complete group is
$O(6)\sim SU(4)$, and the non-anomalous transformations, that are now
represented by $\alpha$ generators, belong to $U(3)\sim SU(3)\times U(1)$.
The anomalous (Goldstone) diquark degrees of freedom belong to the complex
projective space $CP^3=SU(4)/SU(3)\times U(1)$. The same result could be
obtained in a straightforward but tedious way by computing matrix
commutators in an appropriate basis.

As a consistency check we shall demonstrate
that the diquark mass vanishes for zero gluon condensate and zero current
quark masses. We shall also compute the diquark mass and charge radius for
actual value of gluon condensate.

\section{The diquark bosonisation. Gluon condensate as a
source of diquark mass.}

To define the diquark  parameters we should regularize the quark path
integral. We also need a method of extracting a non-invariant part of the
path integral corresponding to anomalous transformations.

To reduce possible regularization dependence [9] we shall use exactly the
same scheme [5] which was developed for chiral bosonisation and generalized
[6] for the presence of diquark variables. Since the parameters of this
scheme were defined through chiral dynamics, we will be able to compare our
results for diquark with pion physics directly.

The basic object is the quark path integral over low scale region
\begin{eqnarray} \label{Z^L}
Z_{\psi}^L &=& (\det\{\hat G \theta(1-(\hat G-M)^2/\Lambda^2)) \})^{1/2}
\nonumber\\
\theta(x) &=& \int_{-\infty}^{\infty} dt\frac{\exp(ixt)}{2\pi i(t-i0)}.
\end{eqnarray}
The parameters $\Lambda$ and $M$ are defined below. The functional
(\ref{Z^L}) can be represented in the form
\begin{eqnarray} 
Z_\psi^L &=& (Z_\psi^L Z_{inv}^{-1}) Z_{inv}
\nonumber\\
Z_{inv}^{-1} &=& \int {\cal D} \Theta (Z_\psi^L(\Theta))^{-1}
\end{eqnarray}
where we integrate over anomalous transformations $\Theta$,
${\cal D} \Theta$ is invariant measure on the corresponding coset space and
$(Z_\psi^L(\Theta))$ is the path integral (\ref{Z^L}) with background
fields transformed as in eq.(\ref{Xi-Theta}). The $Z_{inv}$ does not depend
on degrees of freedom described by $\Theta$. Hence all information over
$\Theta$-noninvariant processes is contained in $(Z_\psi^L Z_{inv}^{-1})$
and the effective action for $\Theta$ can be defined as
\begin{equation} 
(Z_\psi^L Z_{inv}^{-1})=\int {\cal D} \Theta  \exp(iW_{eff}(\Theta))
\end{equation}
The effective action is obtained by integration of the corresponding
anomaly ${\cal A}(x)$
\begin{eqnarray} 
{\cal A}(x;\Theta) &=&
\frac{1}{i}\frac{\delta\ln Z_\psi^L(\Theta)}{\delta\Theta}
\nonumber\\
W_{eff}(\Theta) &=& -\int d^4x \int_{0}^{1} ds {\cal A}(x;s\Theta)\Theta(x)
\end{eqnarray}
Previously [5] this method was applied
to $\pi$-mesons. It was found that the parameters $\Lambda$ and $M$ are
related to the pion decay constant
\begin{equation} 
F_\pi^2=\frac{N_c}{4\pi^2} (\Lambda^2-M^2)
\end{equation}
We will not report here details of computations
 of $W_{eff}(\omega)$ for
the case $\Theta=\omega$, where $\omega$ is given by (\ref{omega}).
They can be performed in same manner as
in the papers [5,6]. Neglecting all external fields except vector gauge
fields
\begin{equation} 
v_\mu=-iQA_\mu+\frac{\lambda^a}{2i}G_\mu^a,
\qquad
Q=
\left(\begin{array}{cc}
2/3 & 0\\
0   & -1/3
\end{array}\right)
\end{equation}
where $A_\mu$ is electromagnetic field, $G_\mu^a$ are gluons
and taking zero current quark masses we
obtain in quadratic order of $\omega$
\begin{eqnarray} \label{W_eff}
\lefteqn{
W_{eff}(\omega)=
\frac{1}{96\pi^2{F_\omega}^2} \tr_{(c,f)} \Biggl\{
6(\Lambda^2-M^2)[D_\mu,\omega^*][D^\mu,\omega]
}
\nonumber\\&&
+[D_\mu,[D^\mu,\omega^*]] [D_\nu,[D^\nu\omega]]
+2[D_\mu,F^{\mu\nu}]
(\omega[D_\nu,\omega^*]+[D_\nu,\omega]\omega^*)
\nonumber\\&&
+(F^{\mu\nu}F_{\mu\nu}\omega\omega^*
-F^{\mu\nu}\omega F_{\mu\nu}^T\omega^*)
\Biggr\}
\end{eqnarray}
where
$[D_\mu,\omega] = (\partial_\mu\omega)+v_\mu\omega+\omega v_\mu^T,
[D_\mu,\omega^*] = (\partial_\mu\omega^*)-v_\mu^T\omega^*-\omega^* v_\mu $
and
$F_{\mu\nu} = (\partial_\mu v_\nu)-(\partial_\nu v_\mu)+[v_\mu,v_\nu]$.
{}From (\ref{W_eff}) we see that the mass of $ud$-diquark
$\omega$ is defined
by the gluon condensate $\langle G_{\mu\nu}^2\rangle$ ($N_c=3$)
\begin{equation} \label{M-ud}
M_\omega^2=-4\pi^2 F_\pi^2
+\sqrt{16\pi^4 F_\pi^4+\frac{\langle G_{\mu\nu}^2\rangle}{12}}
\end{equation}
and vanishes when $\langle G_{\mu\nu}^2\rangle \rightarrow 0$.
For derivation of (\ref{M-ud})
we used
\begin{equation}   
\langle G_{\mu\nu}^a G^{b \mu\nu}\rangle=
\frac{1}{8}\delta^{ab}\langle G_{\mu\nu}^2 \rangle
\end{equation}
For $\langle G_{\mu\nu}^2\rangle = (365 MeV)^4$ we get $M_\omega\approx
300 MeV$. The correction of this evaluation due to quark masses
is provided by $M_\omega^2(m_q\ne 0)=M_\omega^2(m_q=0)+m_\pi^2$.
This gives $M_\omega^2(m_q\ne 0)\approx 340 MeV$, which falls into the
region allowed in the other models [4], though lies close to the lower
boundary. $F_\omega$ is defined by requirement that the residue of the
diquark propagator at $k^2=M_\omega^2$ is unity,
\begin{equation} 
F_\omega^2=\frac{\Lambda^2-M^2}{4\pi^2}+ \frac{1}{12\pi^2}M_\omega^2
\end{equation}

The coefficient before the term
$\partial^2A^\mu
(\omega^*(\partial_\mu\omega)-(\partial_\mu\omega^*)\omega)$
allows us to evaluate the mean
square radius of diquark charge distribution
\begin{equation} 
\langle r^2 \rangle^{1/2} \approx  0.5 fm
\end{equation}
This value is also compatible with other data [4] for diquark effective
radius.

Our desire to describe diquarks as  a quasi-particle similar to
$\pi$ meson has more
than aesthetic grounds. This model allows to explain relatively low mass of
the scalar diquark and include diquark variables in framework of current
algebra and chiral perturbative theory.
As far as we were able to
verify, this suggestion  does not lead to any contradictions. We
obtained quite sensible results for diquark mass and charge radius.
The model has no free parameters. All this indicates that
broken E$\chi$-symmetry deserves further investigations.

This work was supported  by Russian Foundation for Fundamental Studies,
grant 93-02-14378.

\end{document}